\documentclass[runningheads]{llncs}

\usepackage{graphicx}
\usepackage{wrapfig}
\usepackage{amsmath}
\usepackage{amsfonts}
\usepackage{xcolor}
\usepackage{url}

\usepackage{algorithm}
\usepackage{algpseudocode}

\usepackage{lipsum}

\graphicspath{{figures/}}

\begin{document}

\title{Dreamer-CPC: Message Learning with World Models for Decentralized Multi-agent Reinforcement Learning}
\author{
    Taisuke Takayama\inst{1}\orcidID{0009-0001-4127-2091} \and
    Naoto Yoshida\inst{1}\orcidID{0000-0002-9813-0668} \and
    Tadahiro Taniguchi\inst{1,2}\orcidID{0000-0002-5682-2076}
}
\authorrunning{T. Takayama et al.}

\institute{
    Graduate School of Informatics, Kyoto University, Kyoto, Japan \and
    Research Organization of Science and Technology, Ritsumeikan University, Shiga, Japan \\
    \email{takayama.taisuke.73z@st.kyoto-u.ac.jp,\\
        yoshida.naoto.8x@kyoto-u.ac.jp,taniguchi@i.kyoto-u.ac.jp}
}
\maketitle

\begin{abstract}
    In multi-agent reinforcement learning (MARL), inter-agent communication is effective for improving performance under partial observability. Representation learning-based approaches enable decentralized agents to learn messages grounded in their own observations, but they rely only on current observations and cannot convey information accumulated over time. We propose Dreamer-CPC, a decentralized model-based MARL method that integrates message learning based on Collective Predictive Coding (CPC) into the world model of DreamerV3. Each agent independently maintains a world model and a message module, and infers and exchanges messages from the latent states of the world model that reflect the history of past observations and actions. We evaluated Dreamer-CPC in two environments: Observer, a non-cooperative information-sharing task, and CatchApple, a newly introduced task in which task-relevant observations are temporarily missing. In both environments, Dreamer-CPC outperformed IPPO-CPC, an existing CPC-based method that generates messages from current observations, as well as no-communication baselines. In particular, in CatchApple, Dreamer-CPC achieved 4 to 5 times the episode return of IPPO-CPC, demonstrating effective coordination where other methods fail due to missing observations. These results suggest that communication grounded in the latent dynamics of world models can support decentralized decision-making when current observations alone are insufficient.
\end{abstract}

\section{Introduction}

In partially observable and dynamic environments, an agent cannot always acquire sufficient information about task-relevant states from its current local observation alone~\cite{hansen2004dynamic,kaelbling1998planning,nayyar2013decentralized}. This issue is central in decentralized multi-agent settings, where each agent acts based on its own local observation history and task-relevant information may be distributed across agents. Inter-agent communication can mitigate this problem by allowing agents to share information derived from their private observations and observation histories~\cite{albrecht2024multiagent,gronauer2022multiagent,zhu2024survey}. These settings motivate communication mechanisms that convey information accumulated and inferred over time, rather than information contained only in the current local observation.

In multi-agent reinforcement learning (MARL), communication has been shown to enable information sharing among agents and improve performance under partial observability. Many existing methods adopt centralized training with decentralized execution (CTDE), coupling agents through a centralized optimization framework. This includes architectures such as RIAL, DIAL~\cite{foerster2016learning}, and CommNet~\cite{sukhbaatar2016learning}, as well as the use of global value functions~\cite{lowe2017multiagent,sunehag2018value}, which improve policy optimization through gradient propagation across agents~\cite{zhu2024survey}. Parameter sharing among agents is also widely employed~\cite{zhu2024survey}. In addition, many of these methods optimize messages through task rewards and generally assume cooperative settings in which agents learn under a shared reward. While such approaches contribute to engineering performance improvements, they often diverge from the decentralized learning observed in natural agents.

In contrast to these centralized paradigms, representation learning-based approaches enable decentralized agents to learn communication grounded in their own observations. Lin et al.~\cite{lin2021learning} proposed a method in which each agent learns a message representation from its own observation using an autoencoder and broadcasts it to other agents. Meanwhile, Collective Predictive Coding (CPC)~\cite{taniguchi2024collective,taniguchi2025system} from emergent communication research formulates symbol emergence as decentralized Bayesian inference in a joint generative model over agents' observations, providing a theoretical foundation for representation learning-based approaches. Ebara et al.~\cite{ebara2023multiagent} integrated CPC-based message learning into multi-agent reinforcement learning using the Metropolis-Hastings naming game~\cite{taniguchi2023emergent}, demonstrating that discrete messages indicating agents' states can emerge through communication between independent agents without centralized training. MARL-CPC~\cite{yoshida2026reward} formulated autoencoder-based message learning within a variational inference framework and demonstrated that messages useful for coordination can emerge even in non-cooperative settings where rewards are not shared among agents. However, all of these methods generate messages based only on current observations, which is insufficient when task-relevant information is temporarily occluded or when agents must act based on predictions of future states.

We propose Dreamer-CPC, which integrates CPC-based message learning into a world model~\cite{ha2018recurrent} that learns latent dynamics from histories of observations and actions, thereby extending CPC-based message learning beyond current observations. Specifically, each agent learns messages from the latent states of the recurrent state-space model (RSSM)~\cite{hafner2019learning} in DreamerV3~\cite{hafner2025mastering}, enabling communication grounded not only in current observations but also in the history of past observations and actions accumulated by the world model. World model-based MARL methods have been proposed primarily to improve sample efficiency~\cite{egorov2022scalable,wu2023models,xu2022mingling,zhang2025decentralized}, but these methods are all based on the CTDE framework, and the decentralized setting in which each agent learns from its own observation and action history and communication messages alone remains largely unexplored. Nomura et al.~\cite{nomura2025decentralized} proposed a decentralized world model in which each agent maintains an RSSM-based local model and learns messages through InfoNCE loss~\cite{oord2019representation}. However, their policy is learned through behavioral cloning from expert demonstrations and does not involve reinforcement learning in the imagined space of the world model. Dreamer-CPC integrates CPC-based message learning with an RSSM under a unified objective function and establishes a novel model-based MARL framework based on decentralized training with decentralized execution (DTDE) that optimizes policies through imagination rollouts. We evaluated Dreamer-CPC in two environments: Observer~\cite{yoshida2026reward}, a non-cooperative information-sharing task, and CatchApple, a newly introduced environment that requires predictive control under temporally occluded observations.

The contributions of this paper are as follows. First, we propose Dreamer-CPC, a novel deep MARL method that integrates CPC-based communication learning into the latent dynamics of an RSSM, enabling communication learning grounded in the temporal context accumulated by the world model beyond current observations. Second, we show that world model learning and message learning can be jointly optimized under a unified objective function. Third, through experiments in Observer and the newly introduced CatchApple environment, we demonstrated that Dreamer-CPC outperforms IPPO-CPC~\cite{yoshida2026reward}, an existing MARL-CPC method, and no-communication baselines. The results show that communication grounded in the latent dynamics of an RSSM can support coordination when task-relevant observations are temporally unavailable. %
\section{Dreamer-CPC}
\begin{figure}[t]
    \centering
    \includegraphics[width=\linewidth]{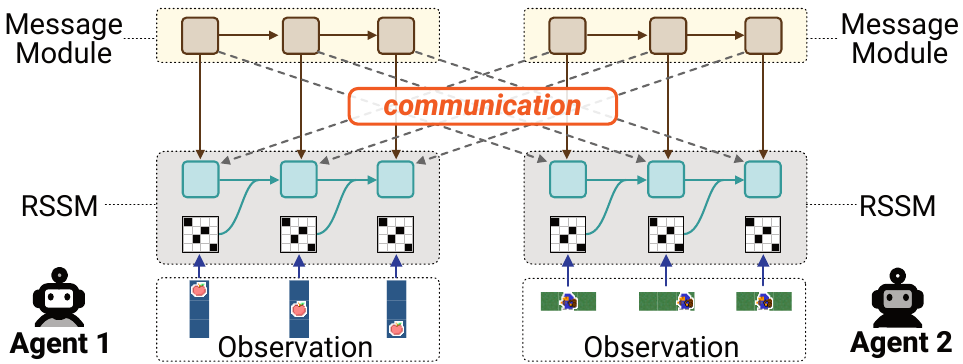}
    \caption{Overview of Dreamer-CPC. Each agent maintains an RSSM-based world model and a message module. Messages are inferred from the latent states of the world model and exchanged between agents.}
    \label{fig:dreamer-cpc_overview}
\end{figure}

We propose Dreamer-CPC, a decentralized model-based MARL method.
Fig.~\ref{fig:dreamer-cpc_overview} provides an overview of Dreamer-CPC.
To enable message learning grounded in temporal latent dynamics beyond
current observations, Dreamer-CPC integrates CPC-based message learning
into the world model and generates messages from its latent states. Each
agent independently maintains a world model and a message module, and
infers and exchanges messages based on the temporal latent dynamics
accumulated by the RSSM.

\subsection{Problem Settings}

We formulate the multi-agent reinforcement learning problem considered in this paper as a partially observable Markov game (POMG)~\cite{hansen2004dynamic}. The game is specified by \(\mathcal{G}=\langle\mathcal{I},\mathcal{S},\{\mathcal{A}^k\}_{k\in\mathcal{I}},\{\mathcal{X}^k\}_{k\in\mathcal{I}},\mu_0,p_{\mathrm{tr}},p_{\mathrm{obs}},\{\rho^k\}_{k\in\mathcal{I}},\gamma\rangle\). Here, \(\mathcal{I}=\{1,\ldots,K\}\) is the set of \(K\) agents, \(\mathcal{S}\) is the state space, \(\mathcal{A}^k\) is the action space of agent \(k\), and \(\mathcal{X}^k\) is the observation space of agent \(k\). The joint action and observation spaces are denoted by \(\mathcal{A}=\prod_{k\in\mathcal{I}}\mathcal{A}^k\) and \(\mathcal{X}=\prod_{k\in\mathcal{I}}\mathcal{X}^k\), respectively. The initial state is sampled as \(s_0\sim\mu_0\), where \(\mu_0\in\Delta(\mathcal{S})\) is the initial state distribution and \(\Delta(\mathcal{S})\) denotes the set of probability distributions over \(\mathcal{S}\). At each time step \(t\), each agent \(k\) selects an action \(a_t^k\in\mathcal{A}^k\) according to its policy \(\pi^k\), using only its available local history. Given the joint action \(a_t=(a_t^1,\ldots,a_t^K)\), the environment state evolves as \(s_{t+1}\sim p_{\mathrm{tr}}(\cdot\mid s_t,a_t)\), and the next joint observation \(x_{t+1}=(x_{t+1}^1,\ldots,x_{t+1}^K)\) is generated as \(x_{t+1}\sim p_{\mathrm{obs}}(\cdot\mid s_{t+1},a_t)\).

Each agent has its own reward function \(\rho^k:\mathcal{S}\times\mathcal{A}\to\mathbb{R}\), and its immediate reward at time step \(t\) is \(r_t^k=\rho^k(s_t,a_t)\). Let \(\pi=(\pi^1,\ldots,\pi^K)\) denote the joint policy, and let \(\pi^{-k}\) denote the policies of all agents except agent \(k\). The discounted return of agent \(k\) is \(G_t^k=\sum_{\ell=0}^{\infty}\gamma^\ell r_{t+\ell}^k\), where \(\gamma\in[0,1)\) is the discount factor. The objective of agent \(k\) is to maximize
\begin{align}
    J^k(\pi^k;\pi^{-k})
    =
    \mathbb{E}_{\mu_0,p_{\mathrm{tr}},p_{\mathrm{obs}},\pi}
    \left[
        G_0^k
        \right],
\end{align}
where the expectation is taken over trajectories induced by the initial state distribution, the transition kernel, the observation kernel, and the joint policy.

\subsection{World Model}\label{sec:world-model}

World models learn the dynamics of the environment from the agent's experience and enable prediction of future observations and rewards given actions. They map observations into a latent space and predict future state transitions within this space. This enables policy learning through imagined trajectories in the latent space without additional interaction with the real environment.

In Dreamer-CPC, each agent $k \in \mathcal{I}$ maintains its own world model, which learns the temporal dynamics of the environment from its local observation and action history as well as messages received from other agents. Each agent's world model is implemented as an RSSM, following DreamerV3. The RSSM has a deterministic recurrent state $h^k_t$ and a stochastic latent state $z^k_t$. At each time step, $h^k_t$ is first computed from the previous latent state $z^k_{t-1}$ and action $a^k_{t-1}$. A prior prediction $\hat{z}^k_t$ is then generated from $h^k_t$ alone, and $z^k_t$ is inferred from $h^k_t$ and the observation $x^k_t$. From $h^k_t$ and $z^k_t$, the observation $x^k_t$, reward $r^k_t$, and continuation flag $c_t \in \{0, 1\}$ are predicted, where $c_t = 1$ indicates that the episode continues beyond time step $t$. The RSSM of agent $k$ is formulated as follows, where $\phi^k$ denotes its parameters:

\vspace{-1em}
\noindent
\begin{minipage}[t]{0.5\linewidth}
    \vspace{0pt}
    \begin{align}
        h^k_t
         & = f_{\phi^k}(h^k_{t-1}, z^k_{t-1}, a^k_{t-1})
        \label{eq:rssm-seq}                              \\
        \hat{z}^k_t
         & \sim p_{\phi^k}(\cdot \mid h^k_t)
        \label{eq:rssm-prior}                            \\
        z^k_t
         & \sim q_{\phi^k}(\cdot \mid h^k_t, x^k_t)
        \label{eq:rssm-post}
    \end{align}
\end{minipage}\hfill%
\begin{minipage}[t]{0.5\linewidth}
    \vspace{0pt}
    \begin{align}
        \hat{x}^k_t
         & \sim p_{\phi^k}(\cdot \mid h^k_t, z^k_t)
        \label{eq:rssm-dec}                          \\
        \hat{r}^k_t
         & \sim p_{\phi^k}(\cdot \mid h^k_t, z^k_t)
        \label{eq:rssm-rew}                          \\
        \hat{c}^k_t
         & \sim p_{\phi^k}(\cdot \mid h^k_t, z^k_t).
        \label{eq:rssm-cont}
    \end{align}
\end{minipage}

\subsection{World Model and Message Learning}

Dreamer-CPC integrates the learning of each agent's RSSM with message learning. In addition to the RSSM, each agent $k$ maintains a recurrent message module. The message module has a message recurrent state $\xi^k_t$, which summarizes the history of previously inferred messages. At each time step, $\xi^k_t$ is updated from the messages $\mathbf{m}_{t-1}$ at the previous step. The message $m^k_t$ is modeled stochastically with a prior distribution (message prior) generated from $\xi^k_t$ alone and a posterior distribution (message posterior) conditioned on $\xi^k_t$ together with the RSSM states $h^k_t$ and $z^k_t$. Denoting the message module parameters by $\psi^k$, the message module is formulated as follows:
\begin{align}
    \xi^k_t     & = g_{\psi^k}(\xi^k_{t-1}, \mathbf{m}_{t-1}) \label{eq:msg-rec}         \\
    \hat{m}^k_t & \sim p_{\psi^k}(\cdot \mid \xi^k_t) \label{eq:msg-prior}               \\
    m^k_t       & \sim q_{\psi^k}(\cdot \mid \xi^k_t, h^k_t, z^k_t). \label{eq:msg-post}
\end{align}
Here, $\hat{\mathbf{m}}_t = (\hat{m}^1_t, \dots, \hat{m}^K_t)$ denotes the joint predicted message collecting the messages sampled from each agent's message prior, and $\mathbf{m}_t = (m^1_t, \dots, m^K_t)$ denotes the joint inferred message collecting the messages sampled from each agent's message posterior.

\begin{figure}[t]
    \centering
    \includegraphics[width=\linewidth]{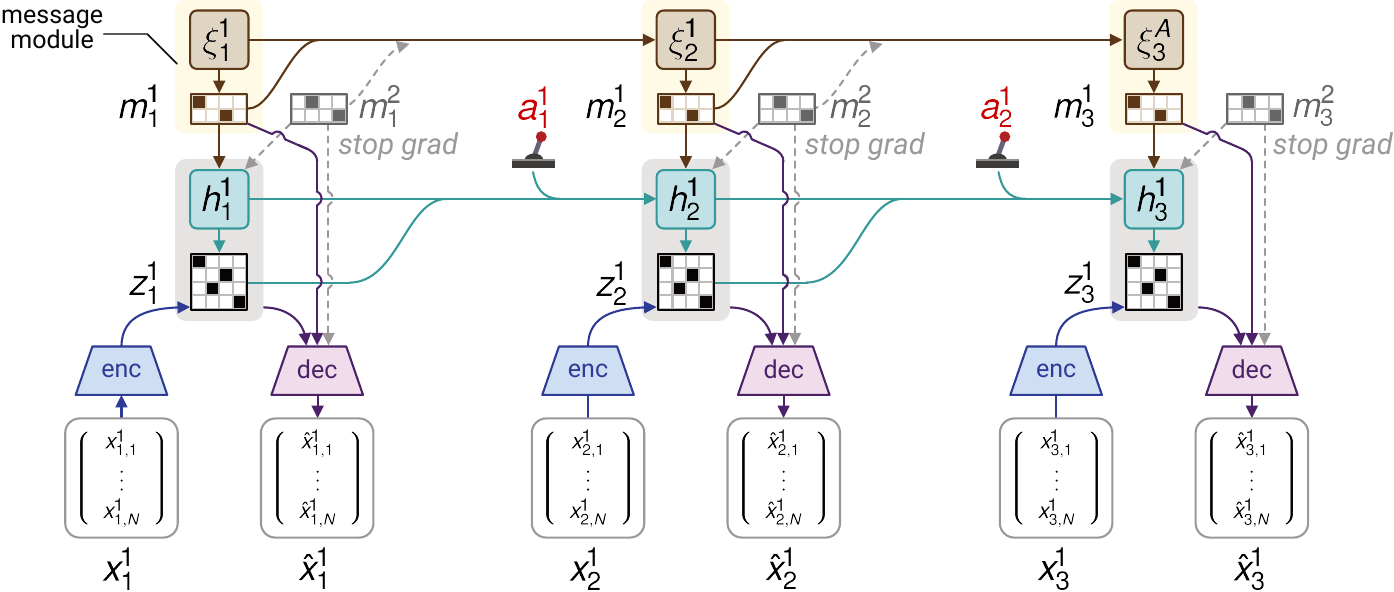}
    \caption{
        Overview of world-model learning in Dreamer-CPC.
        Each agent learns an RSSM state from its local observation history and action history.
        The message module predicts a message from its recurrent state and infers a message from the RSSM state.
        The learned model reconstructs observations using the inferred message and received messages, and predicts rewards and continuation from the RSSM state.
        Messages from other agents are treated with stop gradient, so that communication can be used for prediction without backpropagating through other agents.}
    \label{fig:world-model-learning}
\end{figure}

Messages are incorporated into two components of the RSSM. First, the joint predicted message $\hat{\mathbf{m}}_t$ is fed into the transition of the recurrent state $h^k_t$. Second, the joint inferred message $\mathbf{m}_t$ is added as a conditioning variable for the observation prediction $\hat{x}^k_t$. Accordingly, Eq.~\eqref{eq:rssm-seq} and Eq.~\eqref{eq:rssm-dec} are extended as follows:
\begin{align}
    h^k_t       & = f_{\phi^k}(h^k_{t-1}, z^k_{t-1}, a^k_{t-1}, \hat{\mathbf{m}}_t) \label{eq:msg-rssm-seq} \\
    \hat{x}^k_t & \sim p_{\phi^k}(\cdot \mid h^k_t, z^k_t, \mathbf{m}_t). \label{eq:msg-rssm-dec}
\end{align}
Eq.~\eqref{eq:rssm-prior}, Eq.~\eqref{eq:rssm-post}, Eq.~\eqref{eq:rssm-rew}, and Eq.~\eqref{eq:rssm-cont} remain unchanged from Section~\ref{sec:world-model}. Fig.~\ref{fig:world-model-learning} provides an overview of the learning procedure for this message-augmented world model. To realize decentralized learning, Dreamer-CPC applies stop gradient to messages generated by other agents. This allows each agent to use other agents' messages for prediction without propagating gradients through other agents' parameters.

The world model and the message module are learned jointly by minimizing a unified loss function. We write $p_{\phi^k}(z_t^k \mid h_t^k)$ and $p_{\psi^k}(m_t^k \mid \xi_t^k)$ for the prior densities corresponding to the distributions from which $\hat{z}_t^k$ and $\hat{m}_t^k$ are sampled. For compact notation, we denote the prediction targets of agent $k$ at time step $t$ by $y_t^k=(x_t^k,r_t^k,c_t)$. The prediction likelihood for agent $k$ is factorized as
\begin{align}
    p_{\phi^k}(y_t^k \mid h_t^k,z_t^k,\mathbf{m}_t)
     & =
    p_{\phi^k}(x_t^k \mid h_t^k,z_t^k,\mathbf{m}_t)
    p_{\phi^k}(r_t^k \mid h_t^k,z_t^k)
    p_{\phi^k}(c_t \mid h_t^k,z_t^k).
    \label{eq:prediction-likelihood}
\end{align}
We then define the variational posterior over the stochastic latent state and the message of agent $k$ as
\begin{align}
    q_{\phi^k,\psi^k}(z_{1:T}^k,m_{1:T}^k)
     & =
    \prod_{t=1}^{T}
    q_{\phi^k}(z_t^k \mid h_t^k,x_t^k)
    q_{\psi^k}(m_t^k \mid \xi_t^k,h_t^k,z_t^k).
    \label{eq:joint-variational-posterior}
\end{align}
Assuming fixed initial states, we omit the initial-state terms, and the marginal log-likelihood of the prediction targets can be rewritten as
\begin{align}
     & \log p_{\phi^k,\psi^k}(y_{1:T}^k \mid a_{0:T-1}^k)
    \nonumber
    \\
     & \: =
    \log
    \iint
    \prod_{t=1}^{T}
    \left[
        p_{\phi^k}(y_t^k \mid h_t^k,z_t^k,\mathbf{m}_t)
        p_{\phi^k}(z_t^k \mid h_t^k)
        p_{\psi^k}(m_t^k \mid \xi_t^k)
        \right]
    \, dz_{1:T}^k dm_{1:T}^k
    \label{eq:marginal-likelihood-1}
    \\
     & \: =
    \log
    \mathbb{E}_{q_{\phi^k,\psi^k}}
    \left[
        \prod_{t=1}^{T}
        \left[
            p_{\phi^k}(y_t^k \mid h_t^k,z_t^k,\mathbf{m}_t)
            \frac{
                p_{\phi^k}(z_t^k \mid h_t^k)
            }{
                q_{\phi^k}(z_t^k \mid h_t^k,x_t^k)
            }
            \frac{
                p_{\psi^k}(m_t^k \mid \xi_t^k)
            }{
                q_{\psi^k}(m_t^k \mid \xi_t^k,h_t^k,z_t^k)
            }
            \right]
        \right].
    \label{eq:marginal-likelihood-2}
\end{align}
Applying Jensen's inequality~\cite{jensen1906fonctionsa,kingma2013auto} yields the following variational lower bound:
\begin{align}
    \begin{aligned}
        \mathcal{E}^k
         & =
        \sum_{t=1}^{T}
        \mathbb{E}_{q_{\phi^k,\psi^k}}
        \left[
            \log p_{\phi^k}(y_t^k \mid h_t^k,z_t^k,\mathbf{m}_t)
            \right]
        \\
         & \quad
        -
        \sum_{t=1}^{T}
        \mathbb{E}_{q_{\phi^k,\psi^k}}
        \left[
            D_{\mathrm{KL}}
            \left(
            q_{\phi^k}(z_t^k \mid h_t^k,x_t^k)
            \,\|\,
            p_{\phi^k}(z_t^k \mid h_t^k)
            \right)
            \right]
        \\
         & \quad
        -
        \sum_{t=1}^{T}
        \mathbb{E}_{q_{\phi^k,\psi^k}}
        \left[
            D_{\mathrm{KL}}
            \left(
            q_{\psi^k}(m_t^k \mid \xi_t^k,h_t^k,z_t^k)
            \,\|\,
            p_{\psi^k}(m_t^k \mid \xi_t^k)
            \right)
            \right],
    \end{aligned}
    \label{eq:wm-msg-elbo}
\end{align}
which satisfies
\begin{align}
    \log p_{\phi^k,\psi^k}(y_{1:T}^k \mid a_{0:T-1}^k)
     & \ge
    \mathcal{E}^k.
    \label{eq:wm-msg-elbo-inequality}
\end{align}

In practice, we optimize a stabilized surrogate of the variational objective that balances predictive accuracy with the predictability of latent states and messages. Following DreamerV3, we replace the KL regularizers in Eq.~\eqref{eq:wm-msg-elbo} with KL-balanced losses~\cite{hafner2020mastering} with free bits~\cite{kingma2016improved}. Here, $\beta_{\mathrm{pred}}$, $\beta_{\mathrm{dyn}}$, $\beta_{\mathrm{rep}}$, $\beta_{\mathrm{m\text{-}dyn}}$, and $\beta_{\mathrm{m\text{-}rep}}$ are weighting coefficients for the corresponding loss terms. We set $\beta_{\mathrm{pred}}=\beta_{\mathrm{dyn}}=1$, $\beta_{\mathrm{rep}}=0.1$, $\beta_{\mathrm{m\text{-}dyn}}=1 \times 10^{-3}$, and $\beta_{\mathrm{m\text{-}rep}}=1 \times 10^{-4}$. The resulting objective for agent $k$ is
\begin{align}
     & \mathcal{L}^k_{\mathrm{wm\text{-}msg}} \nonumber \\
     & \: =
    \sum_{t=1}^{T}
    \mathbb{E}
    \left[
    \beta_{\mathrm{pred}}\mathcal{L}_{\mathrm{pred}}^k
    +
    \beta_{\mathrm{dyn}}\mathcal{L}_{\mathrm{dyn}}^k
    +
    \beta_{\mathrm{rep}}\mathcal{L}_{\mathrm{rep}}^k
    +
    \beta_{\mathrm{m\text{-}dyn}}\mathcal{L}_{\mathrm{m\text{-}dyn}}^k
    +
    \beta_{\mathrm{m\text{-}rep}}\mathcal{L}_{\mathrm{m\text{-}rep}}^k
    \right],
    \label{eq:wm-msg-loss}
\end{align}
where
\begin{align}
    \mathcal{L}_{\mathrm{pred}}^k
     & =
    -
    \log p_{\phi^k}
    \left(
    y_t^k
    \mid
    h_t^k,z_t^k,\mathbf{m}_t
    \right)
    \label{eq:wm-msg-pred-loss}
    \\
    \mathcal{L}_{\mathrm{dyn}}^k
     & =
    \max
    \left(
    1,
    D_{\mathrm{KL}}
    \left[
            \operatorname{sg}
            \left(
            q_{\phi^k}(z_t^k \mid h_t^k,x_t^k)
            \right)
            \,\|\,
            p_{\phi^k}(z_t^k \mid h_t^k)
            \right]
    \right)
    \label{eq:wm-dyn-loss}
    \\
    \mathcal{L}_{\mathrm{rep}}^k
     & =
    \max
    \left(
    1,
    D_{\mathrm{KL}}
    \left[
            q_{\phi^k}(z_t^k \mid h_t^k,x_t^k)
            \,\|\,
            \operatorname{sg}
            \left(
            p_{\phi^k}(z_t^k \mid h_t^k)
            \right)
            \right]
    \right)
    \label{eq:wm-rep-loss}
    \\
    \mathcal{L}_{\mathrm{m\text{-}dyn}}^k
     & =
    \max
    \left(
    0.5,
    D_{\mathrm{KL}}
    \left[
            \operatorname{sg}
            \left(
            q_{\psi^k}(m_t^k \mid \xi_t^k,h_t^k,z_t^k)
            \right)
            \,\|\,
            p_{\psi^k}(m_t^k \mid \xi_t^k)
            \right]
    \right)
    \label{eq:msg-dyn-loss}
    \\
    \mathcal{L}_{\mathrm{m\text{-}rep}}^k
     & =
    \max
    \left(
    0.5,
    D_{\mathrm{KL}}
    \left[
            q_{\psi^k}(m_t^k \mid \xi_t^k,h_t^k,z_t^k)
            \,\|\,
            \operatorname{sg}
            \left(
            p_{\psi^k}(m_t^k \mid \xi_t^k)
            \right)
            \right]
    \right).
    \label{eq:msg-rep-loss}
\end{align}

\subsection{Actor--Critic Learning}

Dreamer-CPC follows the same actor--critic learning procedure as DreamerV3. The actor $\pi_{\theta^k}$ of each agent $k \in \mathcal{I}$ parameterizes a distribution over $a^k_t$ conditioned on $h^k_t$ and $z^k_t$, and the critic $v_{\eta^k}$ parameterizes a distribution that estimates the discounted return $G^k_t = \sum_{\ell=0}^{\infty} \gamma^{\ell} r^k_{t+\ell}$ under the same conditioning, where $\theta^k$ and $\eta^k$ denote the actor and critic parameters, respectively. Both are trained with the same objective functions as DreamerV3.

\begin{figure}[t]
    \centering
    \includegraphics[width=\linewidth]{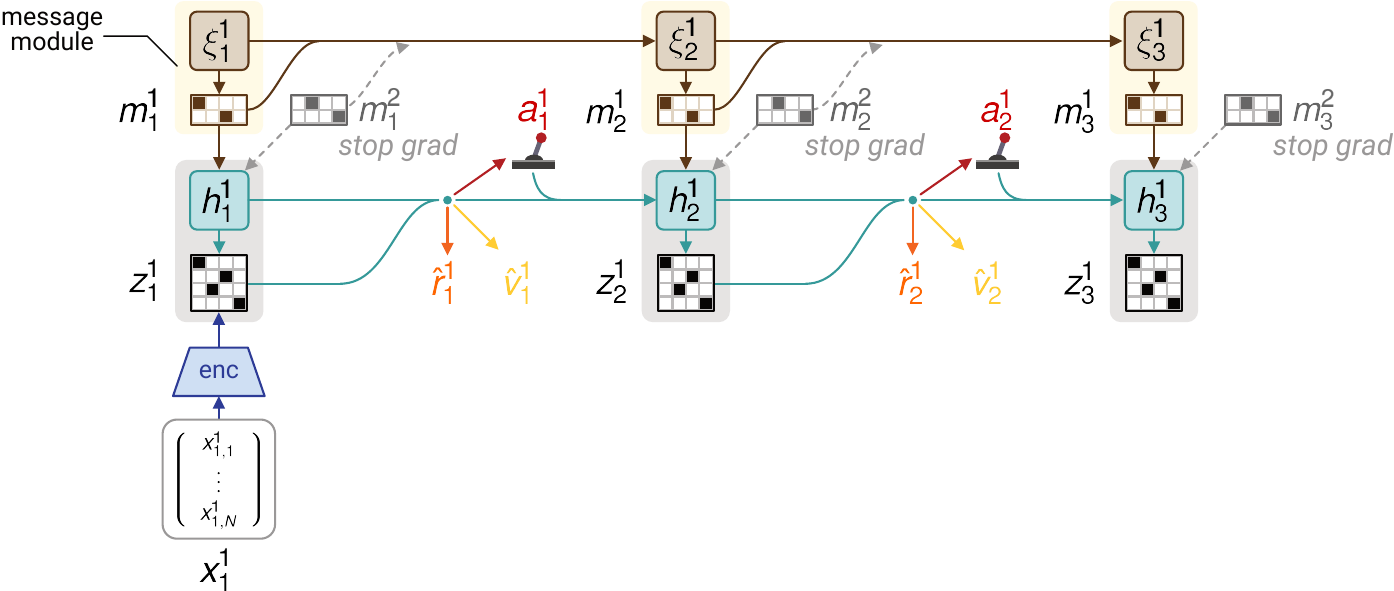}
    \caption{Overview of actor and critic learning with message exchange during imagination. At each imagined step, agents synchronously apply the message generation and inference procedure, share messages, and update their RSSM states using the exchanged messages. Actor and critic losses are then computed on the resulting imagination rollout. Stop gradient labels indicate that received messages are detached during optimization.}
    \label{fig:actor-critic-learning}
\end{figure}

The key difference from DreamerV3 is that imagined trajectories are generated with inter-agent message exchange. Fig.~\ref{fig:actor-critic-learning} illustrates this procedure. At each imagination step, every agent infers a message from its world-model state and exchanges it with other agents. The received messages are then incorporated into the subsequent RSSM transition. Algorithm~\ref{alg:imagination} details this procedure.

\section{Experiments}

We evaluated our proposed method in two MARL environments where each agent receives an individual reward, while information sharing through communication can affect overall performance. In the Observer~\cite{yoshida2026reward} environment, the global state of the environment can be accurately inferred at each time step by aggregating the observations of all agents. In contrast, we constructed the CatchApple environment, in which agents’ observations are temporarily missing at certain time steps. As a result, there are time steps at which the global state cannot be accurately inferred even by aggregating the observations of all agents. Therefore, each agent must infer the current state from its own observation history and the information received from other agents, and act while predicting future states.

We compared Dreamer-CPC with IPPO-CPC~\cite{yoshida2026reward}, IPPO~\cite{dewitt2020independent}, independent DreamerV3 agents without a message channel (no-comm), and DreamerV3 agents that share observations with each other (obs-shared). For IPPO-CPC and Dreamer-CPC, which include a message module, we evaluated two message-channel configurations: a single 20-way one-hot message ($20\times 1$ message) and four independent 5-way one-hot messages ($5\times 4$ messages).

Each agent's RSSM and actor-critic were implemented as approximately 3M-parameter models based on the architecture of DreamerV3. The message module uses a BlockGRU (hidden$=$64, blocks$=$8), and both the message prior and posterior networks are 2-layer MLPs (hidden$=$128) with GELU activation.

\subsection{Observer: Information Sharing without Shared Rewards}
\begin{wrapfigure}[18]{r}{0.35\textwidth}
    \vspace{-6mm}
    \centering
    \includegraphics[width=\linewidth]{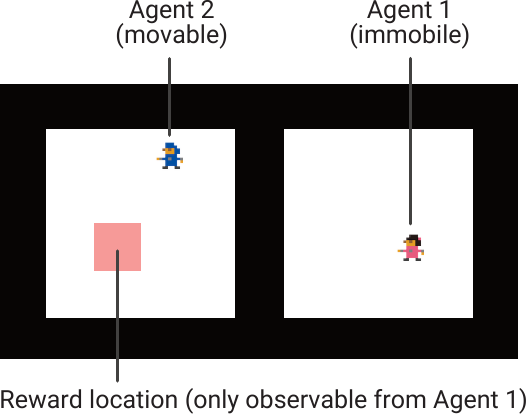}
    \caption{Overview of the Observer environment. Agent 1 observes the reward location, shown as the colored cell, but cannot move, whereas Agent~2 observes only its own grid position and must locate the reward through communication.}
    \label{fig:observer_env}
    \vspace{-3mm}
\end{wrapfigure}
This environment is a non-cooperative MARL environment in which multiple agents receive asymmetric observations, the complete state information of the environment can be reconstructed by integrating their observations, and rewards are not shared, so each agent acts according to its own individual objective. The environment contains two agents. Agent 1 remains stationary and receives no reward, observes a 16-dimensional one-hot vector indicating the reward location, and has a single no-op action, included only to define its action space. In contrast, Agent~2 acts on a $4 \times 4$ grid and observes a one-hot vector indicating its own grid position, but cannot observe the location of the reward.

Agent 2 selects one action from the six actions: $\mathcal{A}_2 = \{\mathtt{up}, \mathtt{down}, \mathtt{left}, \mathtt{right}, \mathtt{still}, \mathtt{dig}\}$.
It can obtain a reward of $+1$ by selecting the $\mathtt{dig}$ action on the cell containing the hidden reward, whereas all other actions incur a penalty of $-0.01$. Episodes were run with a fixed horizon of $H = 200$ steps. Performance is reported as Agent~2's episode return, $G^{2} = \sum_{t=1}^{H} r_{t}^{2}$. Since Agent~1 receives no reward, this environment tests whether message grounding based on representation learning alone can transmit task-relevant information available to another agent.

\subsection{CatchApple: Information Sharing with Temporarily Missing Observations}

\begin{figure}[t]
    \centering
    \includegraphics[width=\linewidth]{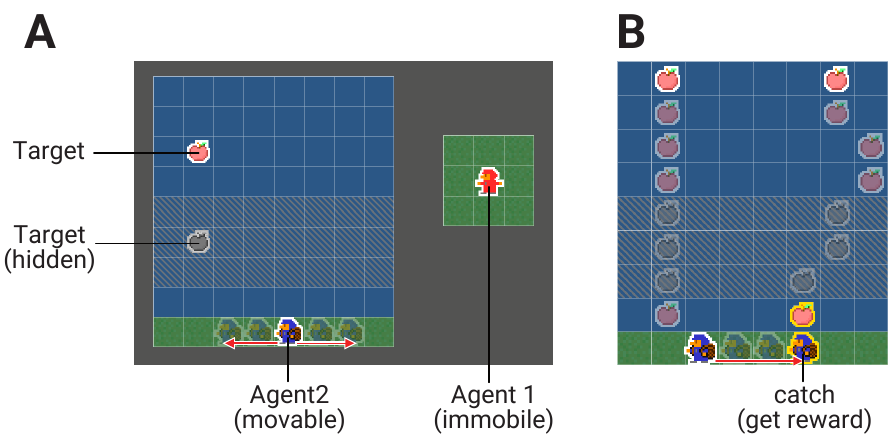}
    \caption{
        Overview of CatchApple environment. (A) The environment contains an $8 \times 8$ falling object area and one ground row. Agent~1 observes the falling object trajectory and receives no reward, whereas Agent~2 controls the catcher on the ground row and observes its own column. The hatched region indicates an occluded region in which the object position is replaced by a hidden flag. (B) The object moves downward, possibly with diagonal motion and boundary reflection, and capture is judged when the object reaches the lowest row of the falling object area.}
    \label{fig:catch-apple-overview}
\end{figure}

This experiment evaluates whether agents can predict the underlying environmental state and act accordingly in an environment where there are time steps at which the full state cannot be reconstructed even by aggregating the observations of all agents. Fig.~\ref{fig:catch-apple-overview} provides an overview of the environment. The environment contains two agents. Agent~1 remains stationary, receives no reward, observes the trajectory of the falling object, and has only a single no-op action. In contrast, Agent~2 controls a catcher located on the ground row and observes its own column, but cannot directly observe the future landing column of the object. Since Agent~2 observes the object's column only after the capture decision has already been made, it cannot capture the current object by reacting to its own observation. Therefore, Agent~2 must predict the object's future landing column from the message received from Agent~1 and move to that column in advance.

The field consists of a falling object area with 8 columns and 8 rows, and a ground row. Agent~1 observes a 65-dimensional one-hot vector representing either the position of the object on the $8 \times 8$ falling object area or a hidden flag indicating that the object is in the occluded row. Agent~1 has only a single no-op action. Agent~2 observes an 8-dimensional one-hot vector indicating its own column and a 9-dimensional one-hot vector indicating either the object's column or a hidden flag. Agent~2 selects one action from three actions, $\mathcal{A}_2 = \{\mathtt{still}, \mathtt{left}, \mathtt{right}\}$.

For each object, the initial column is sampled uniformly at random, and the trajectory type is sampled from three possible types: vertical, diagonally left, and diagonally right. At each step, the object moves downward by one row. For diagonal trajectories, the object additionally moves horizontally by one cell every two rows. When the object reaches a field boundary, its horizontal motion is reflected. A capture is successful if the object's column on the lowest row of the falling object area matches the column of Agent~2. When a capture succeeds, Agent~2 receives a reward of $+1$; at all other time steps, it receives a penalty of $-0.01$. Agent~1 receives no reward. Immediately after the capture judgment, a new object is generated. Episodes were run with a fixed horizon of $H = 500$ steps. Performance is reported as Agent~2's episode return, $G^{2} = \sum_{t=1}^{H} r_{t}^{2}$. As in Observer, Agent~1 receives no reward. This environment further tests whether such message grounding supports coordination under temporally occluded observations.

\begin{figure}[t]
    \centering
    \includegraphics[width=\linewidth]{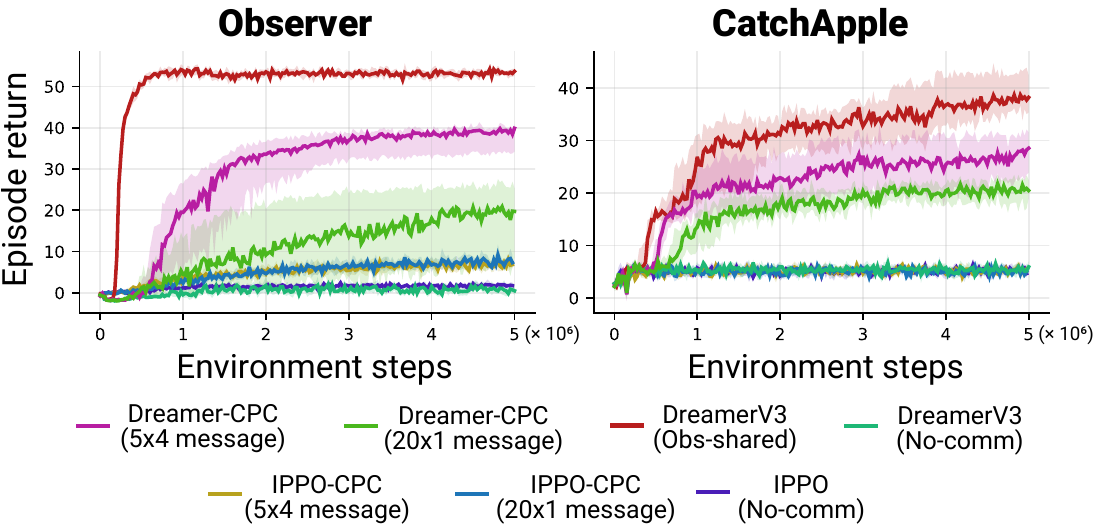}
    \caption{
        Learning curves in Observer and CatchApple.
        Curves show the IQM of episode return over 10 independent runs.
        Shaded regions show 95\% confidence intervals estimated by paired nonparametric bootstrap over runs.
        The left and right panels show Observer and CatchApple, respectively.}
    \label{fig:result}
\end{figure}

\subsection{Results}

Fig.~\ref{fig:result} shows the learning curves in Observer and CatchApple. Each curve represents the interquartile mean (IQM) of the episode return over 10 independent runs, and the shaded regions represent the 95\% confidence intervals estimated by bootstrap resampling of runs with replacement using 2,000 bootstrap samples. In Observer, the obs-shared condition, in which agents share observations, reached an IQM of 53.63. Among the settings without direct observation sharing, Dreamer-CPC achieved the highest performance. The $5 \times 4$ message configuration reached a final-step IQM of 39.85 (95\% CI [34.22, 40.82]), and the $20 \times 1$ message configuration reached 19.72 (95\% CI [8.48, 25.35]). In contrast, IPPO-CPC reached only 7.31 with the $20 \times 1$ message configuration and 6.81 with the $5 \times 4$ message configuration. IPPO, which has no communication channel, and the no-comm DreamerV3 condition remained near zero, with IQMs of 1.74 and 0.48, respectively. Observer is an environment in which the full state can be reconstructed by aggregating the observations of all agents at each time step. These results suggest that Dreamer-CPC can learn information transmission that substantially outperforms the baseline methods, even in non-cooperative settings with severely restricted observability.

In CatchApple, the obs-shared condition again achieved the highest return (IQM 38.21). Dreamer-CPC attained an IQM of 28.53 (95\% CI [23.87, 31.56]) with $5 \times 4$ messages and 20.47 (95\% CI [19.04, 22.45]) with $20 \times 1$ message, substantially outperforming IPPO-CPC ($5 \times 4$: 5.52, $20 \times 1$: 4.98) and the no-comm DreamerV3 condition (5.93). IPPO-CPC, on the other hand, remained at a level comparable to IPPO (4.66), which has no communication channel, and the no-comm condition. In CatchApple, there exist time steps at which observations are temporarily missing, so Agent~2 must act before directly observing the landing column of the object. These results suggest that message learning grounded in the latent dynamics of a world model can contribute to acquiring coordinated behavior in environments where observations are temporarily missing.

\section{Conclusion}

We proposed Dreamer-CPC, a decentralized model-based MARL method that
integrates CPC-based message learning into an RSSM-based world model.
In Dreamer-CPC, each agent learns stochastic messages through a recurrent
message module, and the received messages are used for predicting
observations, rewards, and continuation signals. By treating messages from
other agents as stop-gradient inputs, the method achieves decentralized
learning without introducing direct gradient paths between agents.

Experiments in Observer and CatchApple showed that Dreamer-CPC achieves
higher performance than IPPO-CPC and no-communication baselines in
non-cooperative settings. In particular, the results on CatchApple suggest
that communication learned through a world model can support effective
coordination even when task-relevant observations are temporarily
unavailable.

This work has several limitations. First, we did not analyze what
information the learned messages carry. Qualitative and quantitative
evaluation of the content and structure of messages is necessary for a
deeper understanding of the proposed method. Second, our experiments are
limited to two-agent settings, and validation with a larger number of
agents is needed. Third, both Observer and CatchApple are environments
designed in this paper, and evaluating the effectiveness of Dreamer-CPC
on more general MARL benchmarks remains as future work.
\section*{Acknowledgments}
We would like to thank Prof. Masatoshi Nagano for extensive and valuable advice on the writing of this paper.

\section*{AI Usage Declaration}
We used generative AI tools to assist in the implementation of the proposed method and in the preparation of this manuscript. All generated outputs, including source code and text, were reviewed, verified, and approved by the authors, who take full responsibility for the content of this work.

\newpage
\appendix
\section{Algorithms}\label{sec:algorithms}

\begin{algorithm}[H]
    \caption{World Model and Message Module Learning}
    \label{alg:wm-msg-learning}
    \begin{algorithmic}[1]
        \State Initialize $h^k_0, z^k_0, \xi^k_0, m^k_0$ for all $k \in \mathcal{I}$
        \State Sample trajectories $\{x^k_t, a^k_t, r^k_t, c_t\}_{t=1}^{T}$ from replay buffer for all $k \in \mathcal{I}$
        \For{$t = 1$ \textbf{to} sequence length $T$}
        \For{each agent $k \in \mathcal{I}$}
        \State $\xi^k_t \gets g_{\psi^k}\!\left(\xi^k_{t-1},\, (m^k_{t-1},\, \mathrm{sg}(m^{-k}_{t-1}))\right)$
        \State $\hat{m}^k_t \sim p_{\psi^k}(\cdot \mid \xi^k_t)$
        \EndFor
        \For{each agent $k \in \mathcal{I}$}
        \State $h^k_t \gets f_{\phi^k}\!\left(h^k_{t-1},\, z^k_{t-1},\, a^k_{t-1},\, (\hat{m}^k_t,\, \mathrm{sg}(\hat{m}^{-k}_t))\right)$
        \State $\hat{z}^k_t \sim p_{\phi^k}(\cdot \mid h^k_t)$
        \State $z^k_t \sim q_{\phi^k}(\cdot \mid h^k_t, x^k_t)$
        \State $m^k_t \sim q_{\psi^k}(\cdot \mid \xi^k_t, h^k_t, z^k_t)$
        \EndFor
        \For{each agent $k \in \mathcal{I}$}
        \State $\hat{x}^k_t \sim p_{\phi^k}(\cdot \mid h^k_t, z^k_t, (m^k_t,\, \mathrm{sg}(m^{-k}_t)))$
        \State $\hat{r}^k_t \sim p_{\phi^k}(\cdot \mid h^k_t, z^k_t)$
        \State $\hat{c}^k_t \sim p_{\phi^k}(\cdot \mid h^k_t, z^k_t)$
        \EndFor
        \EndFor
        \State Compute $\mathcal{L}^k_{\mathrm{wm\text{-}msg}}$ and update $\phi^k, \psi^k$ for each agent $k \in \mathcal{I}$
    \end{algorithmic}
\end{algorithm}

\begin{algorithm}[H]
    \caption{Actor--Critic Learning with Imagination Rollouts}
    \label{alg:imagination}
    \begin{algorithmic}[1]
        \State Initialize $h^k_0, \hat{z}^k_0, \xi^k_0, a^k_0, m^k_0$ for all $k \in \mathcal{I}$
        \For{$t = 1$ \textbf{to} imagination horizon $H$}
        \For{each agent $k \in \mathcal{I}$}
        \State $\xi^k_t \gets g_{\psi^k}\!\left(\xi^k_{t-1},\, (m^k_{t-1},\, \mathrm{sg}(m^{-k}_{t-1}))\right)$
        \State $\hat{m}^k_t \sim p_{\psi^k}(\cdot \mid \xi^k_t)$
        \EndFor
        \For{each agent $k \in \mathcal{I}$}
        \State $h^k_t \gets f_{\phi^k}\!\left(h^k_{t-1},\, \hat{z}^k_{t-1},\, a^k_{t-1},\, (\hat{m}^k_t,\, \mathrm{sg}(\hat{m}^{-k}_t))\right)$
        \State $\hat{z}^k_t \sim p_{\phi^k}(\cdot \mid h^k_t)$
        \State $m^k_t \sim q_{\psi^k}(\cdot \mid \xi^k_t, h^k_t, \hat{z}^k_t)$
        \State $a^k_t \sim \pi_{\theta^k}(\cdot \mid h^k_t, \hat{z}^k_t)$
        \State $\hat{r}^k_t \sim p_{\phi^k}(\cdot \mid h^k_t, \hat{z}^k_t)$
        \State $\hat{c}^k_t \sim p_{\phi^k}(\cdot \mid h^k_t, \hat{z}^k_t)$
        \EndFor
        \EndFor
        \State Update actor and critic for each agent following DreamerV3
    \end{algorithmic}
\end{algorithm}
\bibliographystyle{splncs04}
\bibliography{refs}

@book{albrecht2024multiagent,
  author    = {Stefano V. Albrecht and Filippos Christianos and Lukas Sch\"afer},
  title     = {Multi-Agent Reinforcement Learning: Foundations and Modern Approaches},
  publisher = {MIT Press},
  year      = {2024}
}

@article{gronauer2022multiagent,
  title   = {Multi-Agent Deep Reinforcement Learning: A Survey},
  author  = {Gronauer, Sven and Diepold, Klaus},
  year    = {2022},
  journal = {Artificial Intelligence Review},
  volume  = {55},
  number  = {2},
  pages   = {895--943}
}

@article{zhu2024survey,
  title     = {A Survey of Multi-Agent Deep Reinforcement Learning with Communication},
  author    = {Zhu, Changxi and Dastani, Mehdi and Wang, Shihan},
  journal   = {Autonomous Agents and Multi-Agent Systems},
  volume    = {38},
  number    = {1},
  articleno = {4},
  year      = {2024}
}

@inproceedings{sukhbaatar2016learning,
  author    = {Sukhbaatar, Sainbayar and Szlam, Arthur and Fergus, Rob},
  title     = {Learning Multiagent Communication with Backpropagation},
  booktitle = {Advances in Neural Information Processing Systems},
  year      = {2016}
}

@inproceedings{sunehag2018value,
  author    = {Sunehag, Peter and Lever, Guy and Gruslys, Audrunas and Czarnecki, Wojciech Marian and Zambaldi, Vinicius and Jaderberg, Max and Lanctot, Marc and Sonnerat, Nicolas and Leibo, Joel Z. and Tuyls, Karl and Graepel, Thore},
  title     = {Value-Decomposition Networks For Cooperative Multi-Agent Learning Based On Team Reward},
  year      = {2018},
  booktitle = {Proceedings of the 17th International Conference on Autonomous Agents and MultiAgent Systems},
  pages     = {2085--2087}
}

@inproceedings{foerster2016learning,
  author    = {Foerster, Jakob N. and Assael, Yannis M. and de Freitas, Nando and Whiteson, Shimon},
  title     = {Learning to Communicate with Deep Multi-Agent Reinforcement Learning},
  booktitle = {Advances in Neural Information Processing Systems},
  volume    = {29},
  pages     = {2137--2145},
  year      = {2016}
}

@inproceedings{lowe2017multiagent,
  author    = {Lowe, Ryan and Wu, Yi and Tamar, Aviv and Harb, Jean and Abbeel, Pieter and Mordatch, Igor},
  title     = {Multi-Agent Actor-Critic for Mixed Cooperative-Competitive Environments},
  booktitle = {Advances in Neural Information Processing Systems},
  volume    = {30},
  publisher = {Curran Associates, Inc.},
  year      = {2017}
}

@inproceedings{nomura2025decentralized,
  author    = {Nomura, Kentaro and Aoki, Tatsuya and Taniguchi, Tadahiro and Horii, Takato},
  booktitle = {2025 IEEE International Conference on Development and Learning (ICDL)},
  title     = {Decentralized Collective World Model for Emergent Communication and Coordination},
  year      = {2025},
  pages     = {1--8}
}

@article{oord2019representation,
  author  = {van den Oord, Aaron and Li, Yazhe and Vinyals, Oriol},
  title   = {Representation Learning with Contrastive Predictive Coding},
  journal = {arXiv preprint arXiv:1807.03748},
  year    = {2018}
}

@inproceedings{lin2021learning,
  author    = {Toru Lin and
               Jacob Huh and
               Christopher Stauffer and
               Ser{-}Nam Lim and
               Phillip Isola},
  title     = {Learning to Ground Multi-Agent Communication with Autoencoders},
  booktitle = {Advances in Neural Information Processing Systems},
  pages     = {15230--15242},
  year      = {2021}
}

@inproceedings{yoshida2026reward,
  author    = {Yoshida, Naoto and Taniguchi, Tadahiro},
  title     = {Reward-Independent Messaging for Decentralized Multi-Agent Reinforcement Learning},
  booktitle = {Neural Information Processing},
  series    = {Lecture Notes in Computer Science},
  volume    = {16309},
  pages     = {367--382},
  year      = {2026},
  publisher = {Springer}
}

@inproceedings{ebara2023multiagent,
  author    = {Ebara, Hiroto and Nakamura, Tomoaki and Taniguchi, Akira and Taniguchi, Tadahiro},
  title     = {Multi-Agent Reinforcement Learning with Emergent Communication Using Discrete and Indifferentiable Message},
  booktitle = {2023 15th International Congress on Advanced Applied Informatics Winter (IIAI-AAI-Winter)},
  pages     = {366--371},
  year      = {2023}
}

@article{taniguchi2023emergent,
  author  = {Taniguchi, Tadahiro and Yoshida, Yuto and Matsui, Yuta and Hoang, Nguyen Le and Taniguchi, Akira and Hagiwara, Yoshinobu},
  title   = {Emergent Communication through Metropolis-Hastings Naming Game with Deep Generative Models},
  journal = {Advanced Robotics},
  volume  = {37},
  number  = {19},
  pages   = {1266--1282},
  year    = {2023}
}

@article{dewitt2020independent,
  author  = {de Witt, Christian Schr{\"o}der and Gupta, Tarun and Makoviichuk, Denys and Makoviychuk, Viktor and Torr, Philip H. S. and Sun, Mingfei and Whiteson, Shimon},
  title   = {Is Independent Learning All You Need in the {StarCraft} Multi-Agent Challenge?},
  journal = {arXiv preprint arXiv:2011.09533},
  year    = {2020}
}

@inproceedings{hansen2004dynamic,
  author    = {Hansen, Eric A. and Bernstein, Daniel S. and Zilberstein, Shlomo},
  title     = {Dynamic Programming for Partially Observable Stochastic Games},
  booktitle = {Proceedings of the Nineteenth National Conference on Artificial Intelligence},
  pages     = {709--715},
  year      = {2004}
}

@article{nayyar2013decentralized,
  author  = {Nayyar, Ashutosh and Mahajan, Aditya and Teneketzis, Demosthenis},
  journal = {IEEE Transactions on Automatic Control},
  title   = {Decentralized Stochastic Control with Partial History Sharing: A Common Information Approach},
  year    = {2013},
  volume  = {58},
  number  = {7},
  pages   = {1644-1658}
}

@article{kaelbling1998planning,
  title   = {Planning and Acting in Partially Observable Stochastic Domains},
  author  = {Kaelbling, Leslie Pack and Littman, Michael L. and Cassandra, Anthony R.},
  journal = {Artificial Intelligence},
  volume  = {101},
  number  = {1--2},
  pages   = {99--134},
  year    = {1998}
}

@article{taniguchi2025system,
  author  = {Taniguchi, Tadahiro and Hirai, Yasushi and Suzuki, Masahiro and Murata, Shingo and Horii, Takato and Tanaka, Kazutoshi},
  title   = {System 0/1/2/3: Quad-Process Theory for Multitimescale Embodied Collective Cognitive Systems},
  journal = {Artificial Life},
  volume  = {31},
  number  = {4},
  pages   = {465--496},
  year    = {2025}
}

@article{taniguchi2024collective,
  author  = {Taniguchi, Tadahiro},
  title   = {Collective Predictive Coding Hypothesis: Symbol Emergence as Decentralized Bayesian Inference},
  journal = {Frontiers in Robotics and AI},
  volume  = {11},
  year    = {2024}
}

@inproceedings{egorov2022scalable,
  author    = {Vladimir Egorov and
               Alexey Shpilman},
  title     = {Scalable Multi-Agent Model-Based Reinforcement Learning},
  booktitle = {21st International Conference on Autonomous Agents and Multiagent
               Systems, {AAMAS} 2022, Auckland, New Zealand, May 9-13, 2022},
  pages     = {381--390},
  year      = {2022}
}

@inproceedings{xu2022mingling,
  author    = {Zhiwei Xu and
               Dapeng Li and
               Bin Zhang and
               Yuan Zhan and
               Yunpeng Bai and
               Guoliang Fan},
  title     = {Mingling Foresight with Imagination: Model-Based Cooperative Multi-Agent
               Reinforcement Learning},
  booktitle = {Advances in Neural Information Processing Systems},
  year      = {2022}
}

@article{zhang2025decentralized,
  title   = {{Decentralized Transformers with Centralized Aggregation Are Sample-Efficient Multi-Agent World Models}},
  author  = {Zhang, Yang and Bai, Chenjia and Zhao, Bin and Yan, Junchi and Li, Xiu and Li, Xuelong},
  journal = {Transactions on Machine Learning Research},
  year    = {2025}
}

@inproceedings{wu2023models,
  author    = {Wu, Zifan and Yu, Chao and Chen, Chen and Hao, Jianye and Zhuo, Hankz Hankui},
  title     = {Models as Agents: Optimizing Multi-Step Predictions of Interactive Local Models in Model-Based Multi-Agent Reinforcement Learning},
  booktitle = {Proceedings of the AAAI Conference on Artificial Intelligence},
  articleno = {1172},
  numpages  = {9},
  year      = {2023}
}

@article{hafner2025mastering,
  title   = {Mastering Diverse Control Tasks through World Models},
  author  = {Hafner, Danijar and Pasukonis, Jurgis and Ba, Jimmy and Lillicrap, Timothy},
  year    = {2025},
  journal = {Nature},
  volume  = {640},
  number  = {8059},
  pages   = {647--653}
}

@inproceedings{hafner2019learning,
  title     = {Learning Latent Dynamics for Planning from Pixels},
  author    = {Hafner, Danijar and Lillicrap, Timothy and Fischer, Ian and Villegas, Ruben and Ha, David and Lee, Honglak and Davidson, James},
  booktitle = {Proceedings of the 36th International Conference on Machine Learning},
  pages     = {2555--2565},
  year      = {2019}
}

@inproceedings{hafner2020mastering,
  title     = {Mastering {Atari} with Discrete World Models},
  author    = {Hafner, Danijar and Lillicrap, Timothy P. and Norouzi, Mohammad and Ba, Jimmy},
  booktitle = {International Conference on Learning Representations},
  year      = {2021}
}

@inproceedings{ha2018recurrent,
  title     = {Recurrent World Models Facilitate Policy Evolution},
  booktitle = {Advances in Neural Information Processing Systems},
  author    = {Ha, David and Schmidhuber, J{\"u}rgen},
  year      = 2018
}

@inproceedings{kingma2016improved,
  title     = {Improved Variational Inference with Inverse Autoregressive Flow},
  booktitle = {Advances in Neural Information Processing Systems},
  author    = {Kingma, Durk P and Salimans, Tim and Jozefowicz, Rafal and Chen, Xi and Sutskever, Ilya and Welling, Max},
  year      = 2016
}

@inproceedings{kingma2013auto,
  author    = {Diederik P. Kingma and
               Max Welling},
  title     = {Auto-Encoding Variational Bayes},
  booktitle = {2nd International Conference on Learning Representations, {ICLR} 2014,
               Banff, AB, Canada, April 14-16, 2014, Conference Track Proceedings},
  year      = {2014}
}

@article{jensen1906fonctionsa,
  title   = {Sur Les Fonctions Convexes et Les In\'egalit\'es Entre Les Valeurs Moyennes},
  author  = {Jensen, J. L. W. V.},
  year    = {1906},
  journal = {Acta Mathematica},
  volume  = {30},
  number  = {1},
  pages   = {175--193}
}

\end{document}